\documentclass[twocolumn]{article} 
\usepackage{authblk}
\usepackage{fullpage}
\usepackage{graphicx}
\usepackage{xcolor}
\usepackage{algorithm}
\usepackage{algpseudocode}
\usepackage{amssymb}
\usepackage{amsthm}
\usepackage{amsmath}
\usepackage{mathrsfs}
\usepackage[mathcal]{euscript}
\usepackage{bm}
\usepackage{mathabx}
\usepackage[colorlinks,urlcolor=cobalt,citecolor=cobalt,linkcolor=cobalt,pdftex, pdfauthor = Lukas Exl]{hyperref}
\usepackage{tabularx}
\usepackage{booktabs}
\usepackage{setspace} 
\usepackage{wasysym}

\usepackage{ulem}

\definecolor{airforce}{rgb}{0.16,0.32,0.75}
\definecolor{cobalt}{rgb}{0.0,0.28,0.67}

\newtheorem{thm}{Theorem}
\newtheorem{lem}[thm]{Lemma}

\title{\Large{\textbf{Machine learning methods for the prediction of micromagnetic magnetization dynamics}}} 
\author[1,4]{Sebastian Schaffer}
\author[1,4]{Norbert~J.~Mauser}
\author[2,4]{Thomas Schrefl}
\author[3,4]{Dieter Suess}
\author[1,4]{Lukas Exl \thanks{\texttt{lukas.exl@univie.ac.at}}}
\affil[1]{\small Wolfgang Pauli Institute c/o Faculty of Mathematics, University of Vienna, Austria.} 
\affil[2]{Christian Doppler Laboratory for Magnet design through physics informed machine learning, Department of Integrated Sensor Systems, Danube University Krems, Austria}
\affil[3]{Faculty of Physics, University of Vienna, Austria}
\affil[4]{University of Vienna Research Platform MMM Mathematics - Magnetism - Materials, University of Vienna, Austria}

\begin{document}
%
\onecolumn
\maketitle
\date

\noindent\textbf{Abstract.} Machine learning (ML) entered the field of computational micromagnetics only recently. The main objective of these new approaches is the automatization of solutions of parameter-dependent problems 
in micromagnetism such as fast response curve estimation modeled by the Landau-Lifschitz-Gilbert (LLG) equation. Data-driven models for the solution of time- and parameter-dependent partial differential 
equations require high dimensional training data-structures. ML in this case is by no means a straight-forward trivial task, it needs algorithmic and mathematical innovation. Our work introduces theoretical 
and computational conceptions of certain kernel and neural network based dimensionality reduction approaches for efficient prediction of solutions via the notion of low-dimensional feature space integration. 
We introduce efficient treatment of kernel ridge regression and kernel principal component analysis via low-rank approximation. A second line follows neural network (NN) autoencoders as nonlinear data-dependent dimensional 
reduction for the training data with focus on accurate latent space variable description suitable for a feature space integration scheme. We verify and compare numerically by means of a NIST standard problem. 
The low-rank kernel method approach is fast and surprisingly accurate, while the NN scheme can even exceed this level of accuracy at the expense of significantly higher costs. \\  

\noindent\textbf{Keywords.} deep neural networks, nonlinear model order and dimensionality reduction, regularized autoencoders, low-rank approximation, kernel methods, computational micromagnetism.\\

\noindent\textbf{Mathematics Subject Classification.} 	62P35,\, 68T05,\,  65Z05

\twocolumn
\section{Introduction}
\begin{figure}[!t]
\centering 
\includegraphics[scale=1.5]{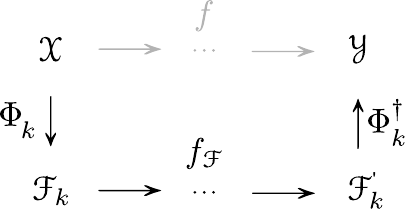}    
\caption{Illustration of the mappings involved in the feature space integration procedure. A map $\Phi_k$ and $\Phi_k^\dagger$ realizes forward and backward transform between original and reduced dimensional space, 
while integration is learned entirely in the lower-dimensional space.}\label{fig:featurespaceint}
\end{figure}

In the recent decades computational micromagnetism has proven useful for simulations guiding design of magnetic devices in applications such as permanent magnets  \cite{fischbacher2018micromagnetics} or magnetic sensors \cite{suess2018topologically}. The theoretical foundation is the continuum theory of micromagnetism, which treats magnetization processes on above quantum mechanical length scales of atoms while still allowing to resolve magnetic domains. The magnetization vector field is modelled as a continuous function inside a magnetic body in three dimensional space. Dynamics of this field in a magnetic material is governed by internal and external fields and mathematically described by the Landau-Lifschitz-Gilbert (LLG) equation, a time-dependent partial differential equation (PDE). This equation is usually treated in finite difference or finite element discretization frameworks \cite{miltat2007numerical,schrefl2007numerical}, where the main computational burden is due to the magnetostatic Maxwell equations posed in whole space     
\cite{abert2013numerical,exl2018magnetostatic}. However, in many applications, such as electronic circuit design and real time process control, the response to a magnetic field needs to be quick. In recent times, data-driven nonlinear reduced order approaches were developed to predict the micromagnetic dynamics depending on the external field \cite{kovacs2019learning,exl2020learning,exl2020prediction} accomplished by machine learning (ML). The common idea is to transform the high-dimensional training magnetization states from simulation results for different field strengths and angles into a feature space where lower-dimensional (approximate) representations exist and then learn the dynamics with respect to the fewer latent variables, see Fig.~\ref{fig:featurespaceint}. 
The main motivation for the kernel based methods introduced in \cite{exl2020learning,exl2020prediction} is to construct a time-stepping predictor on the basis of a non-black-box nonlinear dimensionality reduction with explicit training solutions (in contrast to the extensive optimization needed in deep neural networks). Kernel principal component analysis (kPCA) \cite{scholkopf1997kernel} reduces the feature space dimension, while (kernel ridge) regression models the time-evolution in a $\nu$-step scheme on the level of reduced magnetization representations. 
An effective kernel method realization of the approximate back-transform is trained alongside the kPCA. 
Fig.~\ref{fig:featurespaceint} illustrates the entire dimension-reduced feature space integration scheme, where the time evolution is performed 
on the low-dimensional level in feature space $\mathcal{F}$.
Low-rank variants of all involved kernel methods were introduced in \cite{exl2020prediction} that reduce costs compared to the dense versions and also allow for control of generalization error due to an ”effective rank” while adapting the training sample set. 
The models were tested for the well-established standard problem 4 \cite{mcmichael_mag_nodate} which usually serves as an "elk test" for any novel LLG integration method. The problem is split into two external field cases applied to a magnetic thin film equilibrium s-state and both carefully chosen to trigger a rather complicated dynamics of domain wall motion. 
The overall method performs remarkably well both in terms of accuracy and needed training/prediction times, mainly due to explicitly known solutions for training and prediction. In contrast to this "classical approaches", the authors applied a "modern" convolutional neural network (CNN) to train an autoencoder (-decoder) model that (de-)compresses the magnetization states followed by latent space integration with a feed-forward neural network (FF-NN) \cite{kovacs2019learning}. 
Predictions via the neural network approach exhibited high accuracy in the first field case. However, in the more difficult second region the solution manifold is very nonsmooth and only moderate accuracy could be achieved, even though additional ingredients were incorporated into the encoder/decoder procedure, such as a particular scaling of the z-component of the magnetization and a magnetization length preserving loss term.
Here instead we use a dense autoencoder model with no further constraints that achieves similar or even better results than the (low-rank) kernel method approach.

Our training data consist of computed solutions of the LLG equation for standard problem 4 on a $100 \times 25 \times 1$ nm grid stacked into a \textit{data tensor} $\bar{\mathcal{D}} \in \mathbb{R}^{(s+1)\times n \times (3N+2)}$ defined by
\begin{align}\label{eq:data_tensor}
\bar{\mathcal{D}}(i,:,:) = [\mathbf{h}(t_i)|\mathbf{m}_x(t_i)|\mathbf{m}_y(t_i)|\mathbf{m}_z(t_i)] \in \mathbb{R}^{n \times (3N+2)},
\end{align}
where $\mathbf{m}_q(t_i) \in \mathbb{R}^{n \times N},\, q=x,y,z$ denotes the magnetization component grid vector of length $N$ at equidistantly chosen time points $t_i, \, i=1,\hdots,s,$ for each of the $n$ field values collected in the "tags" $\mathbf{h}(t_i) \in \mathbb{R}^{n \times 2}$ consisting of the randomly chosen external field samples at time $t_i$ with two components for the length $h \in [30,40]\textrm{mT}$ and the angle $\varphi \in [180^\circ,200^\circ]$. We choose $n = 300$ and $s = 100$, giving $m = 30 000$ magnetization samples of dimension $3N = 7500$.

\section{Nonlinear dimensionality reduction}
\subsection{Embedding with autoencoders}

An autoencoder (AE) is an unsupervised model, used to learn descriptive feature space variables, aiming to copy the input to the output. It consists of a encoder part $z = E(x;\boldsymbol{w}_1)$, which maps the input $x$ onto the latent representation $z$ and a decoder $\hat x = D(z;\boldsymbol{w}_2)$, which reconstructs the input from the latent space description. Here $\boldsymbol{w}_1$ and $\boldsymbol{w}_2$ denote the weights. The goal is to minimize the reconstruction error over all $m$ samples
\begin{equation}\label{eq:ae_objective}
	\min_{\boldsymbol{w}_1, \boldsymbol{w}_2}\sum_{i=1}^m \|x_i - \hat x_i\|^2.
\end{equation}
\begin{table}
	\centering
	\begin{tabularx}{\linewidth}{>{\raggedright\arraybackslash}X >{\centering\arraybackslash}X >{\raggedleft\arraybackslash}X}
		\toprule
		Layer & Activation & Output shape \\
		\midrule
		Input & - & $2+ 3N = 7502$  \\
		Dense & elu & $400$ \\
		Dense & elu & $400$ \\
		Dropout & - & $400$ \\
		\midrule
		Dense & linear & $d=40$ \\
		\midrule
		Dense & elu & $400$ \\
		Dropout & - & $400$ \\
		Dense & elu & $400$ \\
		Dense & linear & $3N = 7500$ \\
		\bottomrule
		\medskip
	\end{tabularx}
	\caption{Autoencoder architecture (layer type, activation function and output shape).}\label{tab:ae_arch}
\end{table}

\begin{table}
	\centering
	\begin{tabularx}{\linewidth}{>{\raggedright\arraybackslash}X >{\centering\arraybackslash}X >{\raggedleft\arraybackslash}X}
		\toprule
		Layer & Activation & Output shape \\
		\midrule
		Input & - & $2+\nu d = 202$  \\
		Dense & elu & $400$ \\
		Dense & elu & $300$ \\
		Dense & elu & $300$ \\
		Dense & elu & $200$ \\
		Dense & elu & $100$ \\
		Dense & linear & $d=40$ \\
		\bottomrule
		\medskip
	\end{tabularx}
	\caption{Architecture of the feed-forward neural network used for latent space integration of the autoencoder. For regularization we apply a dropout layer after each non-linear activation.}\label{tab:fs_nn_arch}
\end{table}

\begin{figure}
	\centering
	\includegraphics[width=\linewidth]{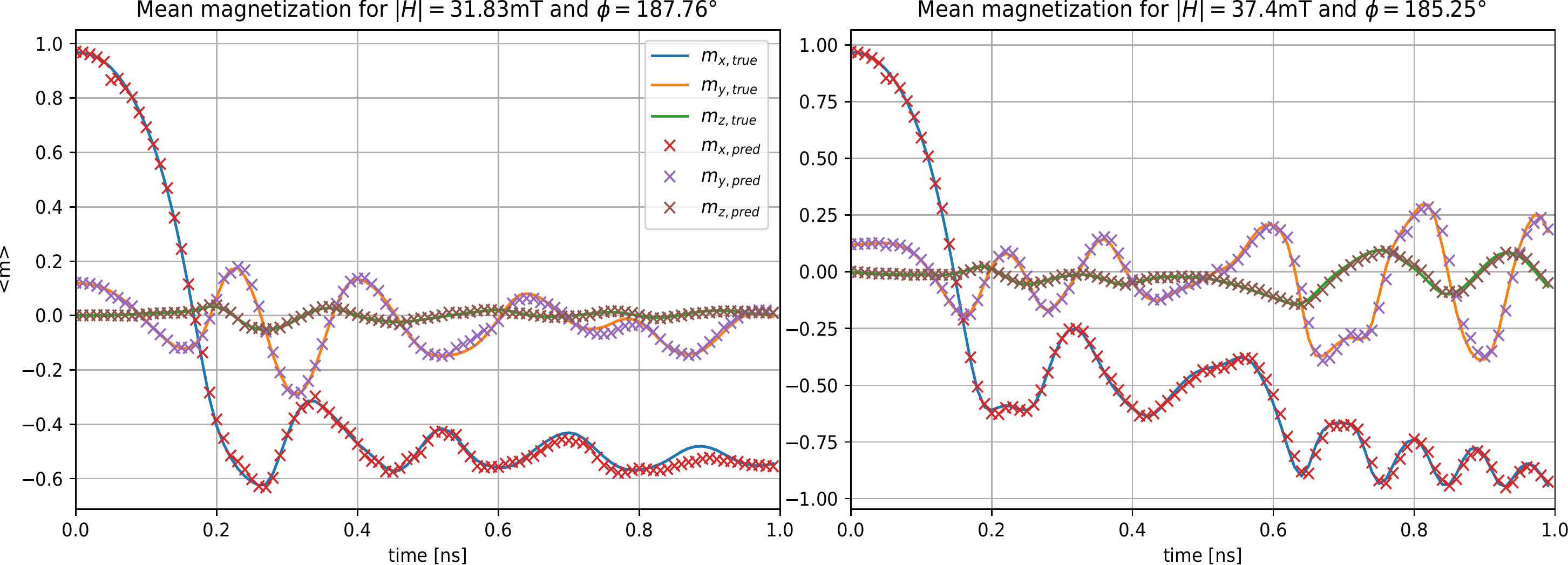}
	\caption{Calculated and predicted mean magnetization for two different magnetic field domains using the autoencoder model and FF-NN for feature-space integration with $d=40$, $\nu=5$ and $f_t=15$. Architectures from Table \ref{tab:ae_arch} and \ref{tab:fs_nn_arch}.}
	\label{fig:ae_mean_mag}
\end{figure}

\begin{figure}
	\centering
	\includegraphics[width=\linewidth]{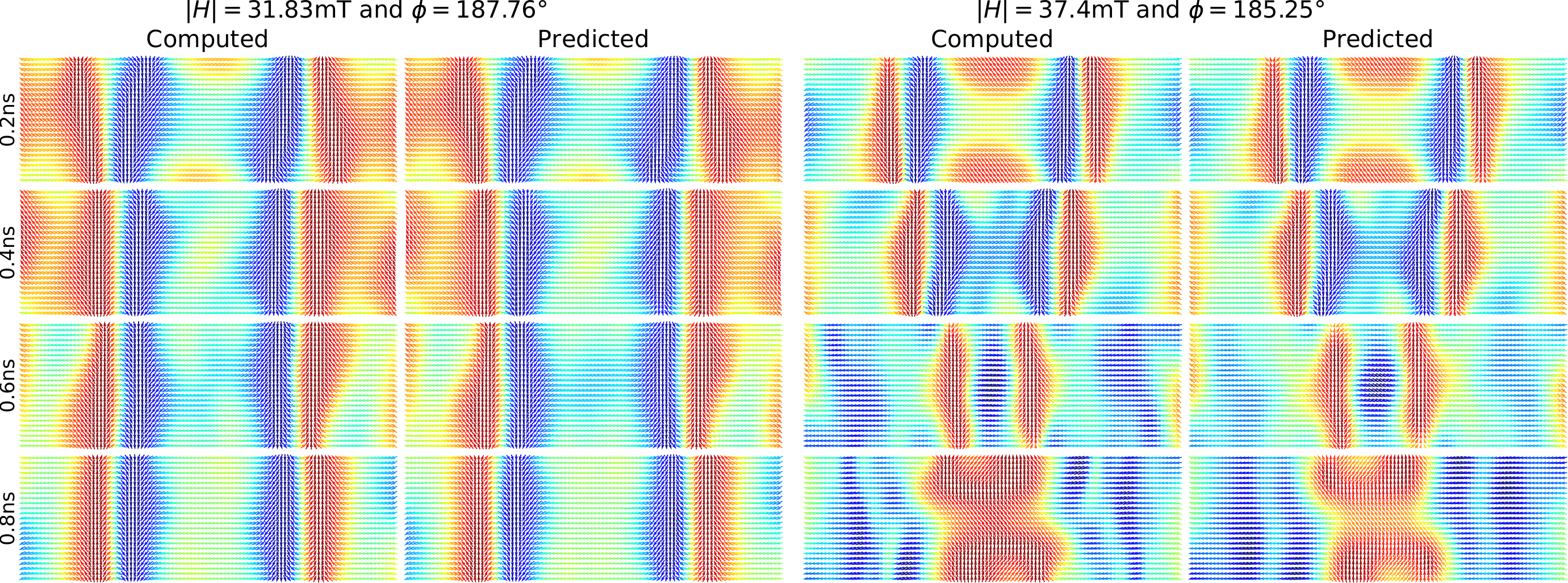}
	\caption{Calculated and predicted magnetization states for four discrete time steps and two different magnetic field domains using the autoencoder model and FF-NN for feature-space integration with $d=40$, $\nu=5$ and $f_t=15$. Architectures from Table \ref{tab:ae_arch} and \ref{tab:fs_nn_arch}.}
	\label{fig:ae_mag_states}
\end{figure}

\begin{figure}
	\centering
	\includegraphics[width=\linewidth]{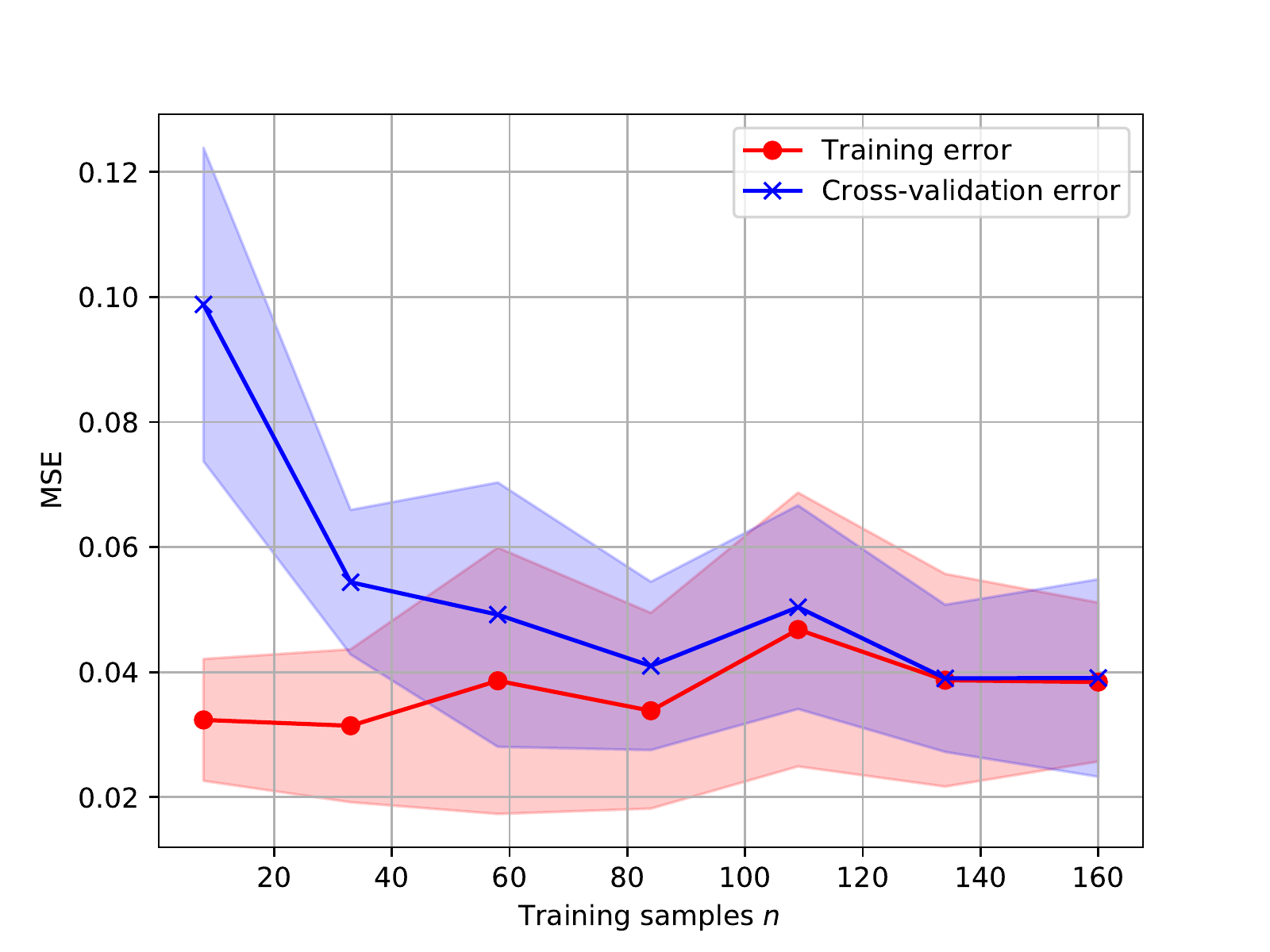}
	\caption{Learning curve based on a $10$-fold cross-validation with a testset size of $20\%$, where the prediction error is averaged over all timesteps. The parameters were chosen as follows: 
	sample size $n = 200$, autoencoder learning rate $l_{AE} = 0.0002$ and FFNN learning rate $l_{FFNN} = 0.0002$.}\label{fig:learning_curve_ae}
\end{figure}

In our model, the encoder and the decoder take the form of a FF-NN. The architecture can be seen in Table \ref{tab:ae_arch}. We apply a minor change to the original structure, by also using the magnetic field as extra input, but it is not included in the output. The AE has a very simple structure, and indeed we observed that already only one hidden dense layer for the encoder/decoder would give acceptable results. We use the \textit{Adam} optimizer \cite{Kingma2014adam} for training the models, with a low learning rate of $2\mathrm{e}{-4}$. Training the AE is straightforward and the main difference to \cite{kovacs2019learning} is the use of a dense AE, without any norm-preserving objective term. A more difficult task is the training of the NN used for feature space integration, since a simple time-stepping scheme will easily overfit, we use a forward looking objective, as applied in \cite{kovacs2019learning, kim2019deepFluids} to gain accuracy.
Let $c_j^i$ be the low dimensional embedding of $\mathcal{\bar D}(i, j, :)$ for time $i$ and field sample $j$.
The NN-predictor model with weights $\boldsymbol{w}$ is
\begin{equation}\label{eq:nn_predictor}
\mathcal{N} : \mathbb{R}^{2 + \nu d} \rightarrow \mathbb{R}^d, \, \mathcal{N}([\hat c_j^{i-\nu+1},\dots, \hat c_j^i]; \boldsymbol w, h_j(t_i)) \mapsto \hat c_j^{i+1},
\end{equation}
where $\hat c_j^i$ is an approximation of $c_j^i$ for $i > \nu\in\mathbb{N}$ and for discrete times $i \leq \nu$ we set $\hat c_j^i  = c_j^i$, compare Table \ref{tab:fs_nn_arch}. Similar to equation \eqref{eq:data_tensor}, $\mathcal{C}(i,:,:)= [c_1^i, \dots, c_n^i]^T \in \mathbb{R}^{n \times d} $ defines the latent space description of the tensor $\bar{\mathcal{D}}(i,:,:)$ at time step $i$ and for all $i > \nu$, $\mathcal{\hat C}(i,:,:) = [\hat c_1^i, \dots, \hat c_n^i]^T \in \mathbb{R}^{n \times d}$ is the prediction of $\mathcal{C}(i,:,:)$ using the predictor in \eqref{eq:nn_predictor} for each field sample. Now we minimize the objective function
\begin{equation}\label{eq:fs_nn_objective}
\min_{\boldsymbol w}\sum_{j=1}^{f_t} \sum_{i=\nu + j}^s \|\mathcal{C}(i,:,:) -  \mathcal{\hat C}(i,:,:)\|^2_F.
\end{equation}
In our model we set $\nu=5$ and $f_t=15$ and train the NN to predict 15 future time steps accurately. Fig. \ref{fig:ae_mean_mag} and \ref{fig:ae_mag_states} show mean magnetization and magnetization snapshots 
for two randomly chosen fields not contained in the training set. Fig.~\ref{fig:learning_curve_ae} shows the learning curve based on a $10$-fold cross-validation with a testset size of $20\%$, where 
the prediction error is averaged over all timesteps.   
Implementation was done in keras \cite{chollet2015keras}.

\subsection{Embedding with kernel methods}\label{kernels}
Alternatively, we describe an unsupervised learning approach with kernels which encapsulate the high dimensional input magnetization. 
Suppose we have $m\in\mathbb{N}$ given sample data points from a set $\mathcal{X}$, i.e., $x_k \in \mathcal{X},\,k=1,\hdots,m$ with the associated  \textit{sample (centered) co-variance matrix} $C := \frac{1}{m} \sum_{j=1}^m(x_j - \langle x \rangle)(x_j - \langle x \rangle)^T$.
The eigenvalue problem related to $C$ reveals the \textit{principal axes} as eigenvectors which form an orthonormal system and where the amount of variance along these axes is given by the respective eigenvalues. The coordinates in the eigensystem are the \textit{principal components}.  
The system of eigenvectors associated with the largest $d$ eigenvalues encompasses the maximal possible amount of variance under all orthogonal systems of dimension $d$. It can be very insightful to perform an orthogonal (basis) transformation into the above eigensystem, known as (linear) \textit{Principal Component Analysis (PCA)}. In fact, if only the largest principal components are kept, most information comprised by the given data will be conserved in the transformed but lower-dimensional system. New data points drawn from the same underlying distribution as the original data can now be transformed to the \textit{trained} PCA system. In this sense (linear) PCA is an \textit{unsupervised or manifold learning} method useful to detect most relevant structure in data. However, the linearity comes with limitations, that is, often the data live on a (nonlinear) manifold where it might be beneficial to apply nonlinear maps on the data before the unsupervised learning approach. Computationally more important, if we assume the data to be discrete magnetization configurations from an computational grid, the sample co-variance matrix approach will be too expensive. On the other hand kernel functions can be used to encapsulate the samples leading to a nonlinear reformulation of PCA in its dual form to work with a gram matrix instead \cite{scholkopf1997kernel,lee2007nonlinear}. We use a so-called positive definite kernel function for that purpose, such as the \textit{(Gaussian) radial basis function (RBF)} 
\begin{align}\label{eqn:rbf}
 k:\,\mathcal{X}\times \mathcal{X} \rightarrow \mathbb{R}^+,\, k(x,y) = e^{-\gamma \|x-y\|^2}
\end{align}
which encapsulates the samples in an associated gram matrix $K[\mathbf{x}] \in \mathbb{R}^{m \times m}$ 
 defined as $K_{ij} = k(x_i,x_j),\,i,j = 1,\hdots,m$. A kernel not only encapsulates the data from $\mathcal{X}$ it  also defines a kind of \textit{similarity measure} through an inner product in a certain Hilbert space. In fact, a kernel $k$ is associated with a (nonlinear) \textit{feature space map} $\phi_k:\,\mathcal{X} \rightarrow \mathcal{F}_k$ which embeds the data from $\mathcal{X}$ into the \textit{feature space} $\mathcal{F}_k$, see e.g. \cite{saitoh1988theory} for the theoretical background. The key is that the inner product in $\mathcal{F}_k$ can be expressed by the kernel $k$ only, that is
\begin{align}\label{eqn:kerneltrick}
 \phi_k(x) \cdot \phi_k(y) = k(x,y),
\end{align} 
which is known as the \textit{kernel trick}. Hence, it is possible to extend linear structural analysis for data, like the (linear) PCA, to their nonlinear kernel analogues by substituting the inner products by kernel evaluation. 
Note also the local isometry of the RBF kernel, a property that is usually desired when constructing data-dependent kernels \cite{weinberger2004learning}. In fact, for two data points $x,y \in \mathcal{X}$ the distance metric induced by \eqref{eqn:rbf} is $\|\phi_k(x)-\phi_k(y)\|_{\mathcal{F}_k}^2 = 2(1-e^{-\gamma \|x-y\|^2}) $ which gets locally Euclidean for small distances, i.e., 
\begin{align}
 \|\phi_k(x)-\phi_k(y)\|_{\mathcal{F}_k}^2 \approx 2\gamma \,\|x-y\|^2. 
\end{align}
 Similarly for angles: We have $\cos{\hat{\psi}} = \phi_k(x)\cdot \phi_k(y) = e^{-\gamma \|x-y\|^2}$ with $\hat{\psi}$ denoting the angle between the mapped points $\phi_k(x)$ and $\phi_k(y)$, which are always normalized to unity. Now, assume normalized data points $x,y \in \mathcal{X}$, then $\|x-y\|^2 = \|x\|^2+\|y\|^2-2\,x\cdot y =  2-2\cos{\psi}$ with $\psi$ 
being the angle between $x$ and $y$, and hence, 
for small angles (via Taylor expansion) there again holds approximately a linear relation 
\begin{align}
 \hat{\psi}^2 \approx  2 \gamma \,\psi^2.
\end{align}
However, a great advantage of the RBF kernel is its fast decay for larger distances in $\mathcal{X}$ resulting in small cosines (i.e. projections) in $\mathcal{F}_k$, that is, $\cos{\hat{\psi}} = k(x,y) = e^{-\gamma \|x-y\|^2}$ is small, and hence a rapidly decaying spectrum of the associated kernel matrix \cite{williams2000effect,wathen2015spectral}. This is now key to the low-rank approach in \cite{exl2020prediction} which allows to \text{learn} a realization of the feature space map $\phi_k$ in the sense of a low-rank decomposition of the kernel matrix. As a consequence, computations simplify and get cheaper.  Hence, we can efficiently learn structure using the mapped data in feature space with a \textit{kernelized} version of linear PCA, the \textit{low-rank kernel principal component analysis} ($\ell$-kPCA) \cite{exl2020prediction,scholkopf1997kernel}:

%
For large sample size $m$ the computation of the Gram matrix and the associated eigenvalue problems in the kPCA might be too expensive or intractable. 
A way to overcome this burden is to learn a low-rank approximation of the Gram matrix 
\begin{align}
K[\mathbf{x}]\approx \Phi_r\,{\Phi_r}^T,
\end{align}
with $\Phi_r := \Phi_r[\mathbf{x}] = K[\mathbf{x}_m,\mathbf{x}_r]\,K[\mathbf{x}_r]^{-1/2} \in \mathbb{R}^{m \times r}$ by randomly choosing a training subset $\mathbf{x}_r \subseteq \mathbf{x}:=\{x_1,\hdots,x_m\},\, r\leq m$ of the given data \cite{williams2001using,fowlkes2004spectral}. This way, we achieve a computational realization of the feature space map that can be used to simplify computations in the kPCA, its approximate back transform and kernel ridge regression. 
The key advantage of the above low-rank decomposition of the Gram matrix is the direct use of the feature vectors, which is normally not possible due to lack of knowledge of the feature space map. One can use the learned map to predict the $\phi_k$-images of new (unseen) data, as well as 
solve the eigenvalue problems in the kPCA associated with the kernel matrix in an efficient way.  
Further information concerning the required storage and cost, the solution of the low-rank eigenvalue problem as well as numerical validation of $\ell$-kPCA can be found in \cite{exl2020prediction}.\\
The mapped data need to be back-transformed into original space $\mathcal{X}$, which is the non-trivial task of (approximate) pre-image computation. We do so by learning a pre-image map while establishing the $\ell$-kPCA. Assume two inner product spaces $\mathcal{X}$ and $\mathcal{Y}$ where we want to model a dependency of input data $\mathbf{x} = \{x_1,\hdots,x_m\} \subseteq \mathcal{X}$ and 
output data $\tilde{\mathbf{x}} = \{\tilde{x}_1,\hdots,\tilde{x}_m\} \subseteq \mathcal{Y}$ by a linear map. Kernel ridge regression (kRR) estimates a linear map $W:\, \mathcal{F}_k \rightarrow \mathcal{Y}$ fulfilling \cite{shalev2014understanding, welling2013kernel}
\begin{align}\label{eqn:krr}
 \min_W \frac{1}{2} \sum_{i=1}^m \|\tilde{x}_i - W\cdot\phi_k (x_i) \|^2 + \frac{\alpha}{2} \|W\|^2,
\end{align}
where $\alpha >0$ is a regularization parameter. The (dual) solution is
\begin{align}\label{eqn:dualsolkrr}
 W^T &\,= 
 \,\phi_k(X)^T\cdot \big(K[\mathbf{x}] + \alpha I \big)^{-1} \tilde{X},
\end{align}
where we assembled the data $\mathbf{x}$ and $\tilde{\mathbf{x}}$ into arrays $X$ and $\tilde{X}$ of shape $m \times \textrm{dim}(\mathcal{X})$ and $m \times \textrm{dim}(\mathcal{Y})$, respectively, and $\phi_k$ is assumed to act on 
the rows of $X$ with $\phi_k(X)$ being of size $m \times \textrm{dim}(\mathcal{F}_k)$. $W^T$ in \eqref{eqn:dualsolkrr} is of size $\textrm{dim}(\mathcal{F}_k) \times \textrm{dim}(\mathcal{Y})$.
Once the kRR training phase \eqref{eqn:dualsolkrr} has been completed, the kRR estimator applied on new data $\mathbf{y} = \{y_1,\hdots,y_\ell\} \subseteq \mathcal{X}$ assembled in $Y$ of shape $\ell \times \textrm{dim}(\mathcal{X})$ takes the form 
\begin{align}\label{eqn:kRRest}
\phi_k(Y)\cdot W^T = K[\mathbf{y},\mathbf{x}] \big(K[\mathbf{x}] + \alpha I \big)^{-1} \tilde{X},
\end{align}
of size $\ell \times \textrm{dim}(\mathcal{Y})$.
Using a low-rank approximation of the kernel matrix $K[\mathbf{x}] \approx \Phi_r[\mathbf{x}]\Phi_r[\mathbf{x}]^T$ with $\Phi_r[\mathbf{x}]\in \mathbb{R}^{m \times r},\,m\geq r$ and $K[\mathbf{y},\mathbf{x}] \approx \Phi_r[\mathbf{y}]\Phi_r[\mathbf{x}]^T$ with the predicted 
$\Phi_r[\mathbf{y}] \in \mathbb{R}^{\ell \times r}$, we can turn \eqref{eqn:kRRest} into
\begin{align}\label{eqn:kRRest_lowrank}
\phi_k(Y)\cdot W^T \approx \Phi_r[\mathbf{y}]\Phi_r[\mathbf{x}]^T \big(\Phi_r[\mathbf{x}]\Phi_r[\mathbf{x}]^T + \alpha I \big)^{-1} \tilde{X}.
\end{align}
We can use the \textit{push-through identity} to get the following computationally more convenient form.
\begin{lem}[Low-rank kRR ($\ell$-kRR)]\label{lem:lowrank_kRR}
 Given training data $\mathbf{x} = \{x_1,\hdots,x_m\}\subseteq \mathcal{X}$ and a low-rank approximation of the Gram matrix $K[\mathbf{x}] \approx \Phi_r[\mathbf{x}]\Phi_r[\mathbf{x}]^T$ with $\Phi_r[\mathbf{x}]\in \mathbb{R}^{m \times r}$ for $m \geq r$. 
 Further let $\mathbf{y} = \{y_1,\hdots,y_\ell\}\subseteq \mathcal{X}$. Then the kRR estimator \eqref{eqn:kRRest} from problem \eqref{eqn:krr} gets
\begin{align}\label{eqn:kRRest_lowrank2}
 \phi_k(Y)\cdot W^T  \approx \Phi_r[\mathbf{y}]\, (\Phi_r[\mathbf{x}]^T\, \Phi_r[\mathbf{x}] + \alpha I)^{-1}\Phi_r[\mathbf{x}]^T \tilde{X},
\end{align}
with $\Phi_r[\mathbf{y}] \in \mathbb{R}^{\ell \times r}$. 
\end{lem}

\textit{Proof.}\quad 
 From the identity $A(BA+\alpha I) = (AB+\alpha I)A$ for appropriately sized matrices $A$ and $B$ we get $A(BA+\alpha I )^{-1} = (AB+\alpha I)^{-1}A$ by left multiplication with $(AB+\alpha I)^{-1}$ and right multiplication with $(BA+\alpha I)^{-1}$, where we assume the existence of the respective inverse matrices.  
  Thus we have $\Phi_r[\mathbf{x}]^T \big(\Phi_r[\mathbf{x}]\Phi_r[\mathbf{x}]^T + \alpha I \big)^{-1} = (\Phi_r[\mathbf{x}]^T\, \Phi_r[\mathbf{x}] + \alpha I)^{-1}\Phi_r[\mathbf{x}]^T$, and therefore 
  \eqref{eqn:kRRest_lowrank2} follows from \eqref{eqn:kRRest_lowrank}. 
\hfill $\Square$\\

\begin{figure}
	\centering
	\includegraphics[width=\linewidth]{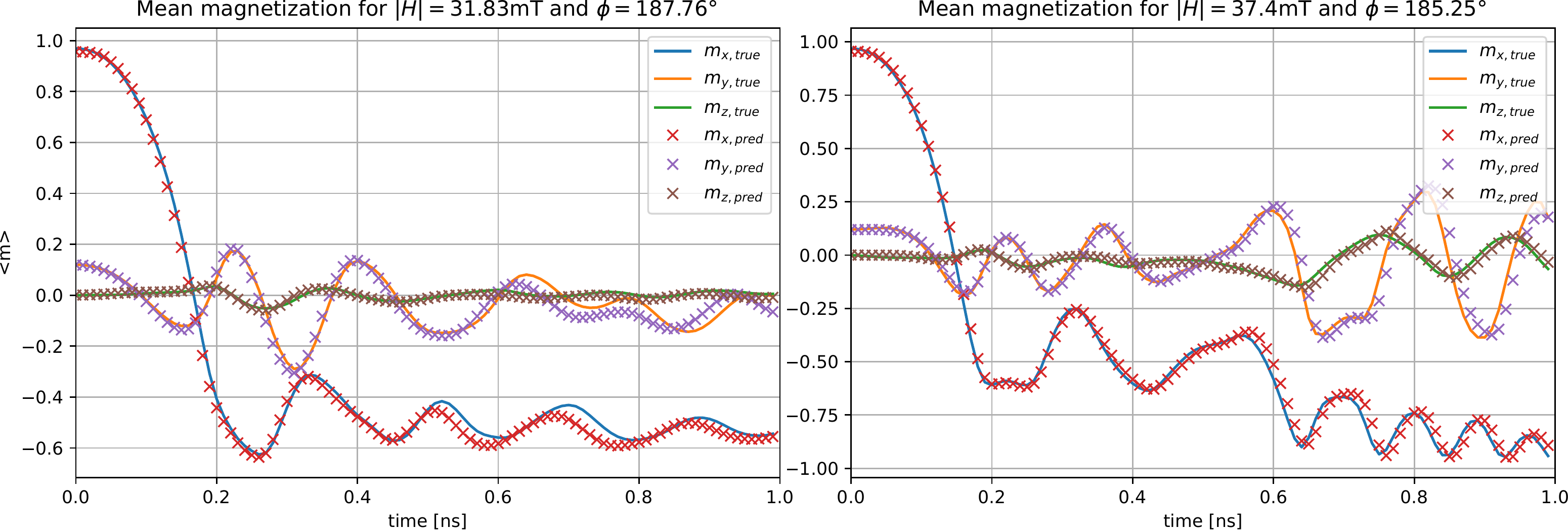}
	\caption{Calculated and predicted mean magnetization for two different magnetic field domains using the $\ell$-kPCA embedding and $\ell$-kRR for feature space integration with low-rank $50$ (compared to $n=300$). We use a RBF with $\gamma=1/(3N)$ and $40$ principal components for the $\ell$-kPCA. For the $\ell$-kRR we also use a RBF with $\gamma=1$ and $\nu=5$ time steps.}
	\label{fig:kpca_mean_mag}
\end{figure}

\begin{figure}
	\centering
	\includegraphics[width=\linewidth]{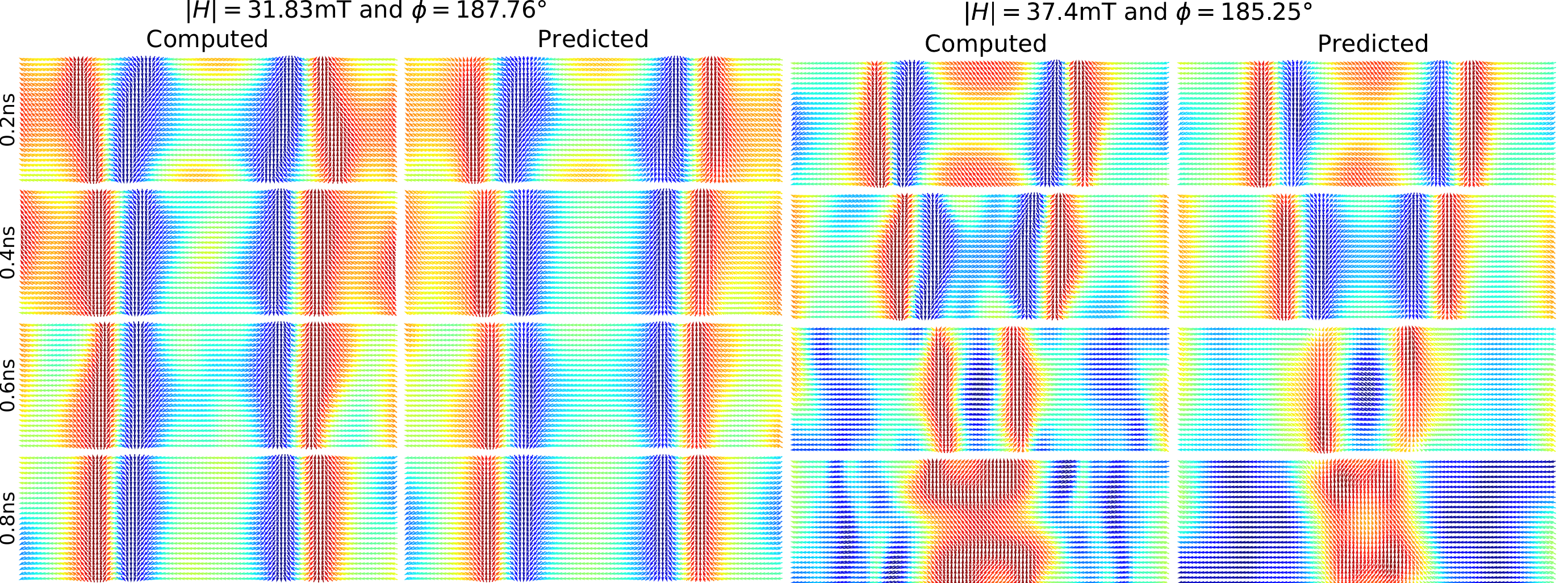}
	\caption{Calculated and predicted magnetization states for four discrete time steps and two different magnetic field domains using the $\ell$-kPCA embedding and $\ell$-kRR for feature space integration with low-rank $50$ (compared to $n=300$). We use a RBF with $\gamma=1/(3N)$ and $40$ principal components for the $\ell$-kPCA. For the $\ell$-kRR we also use a RBF with $\gamma=1$ and $\nu=5$ time steps.}
	\label{fig:kpca_mag_states}
\end{figure} 

\begin{figure}
	\centering
	\includegraphics[width=\linewidth]{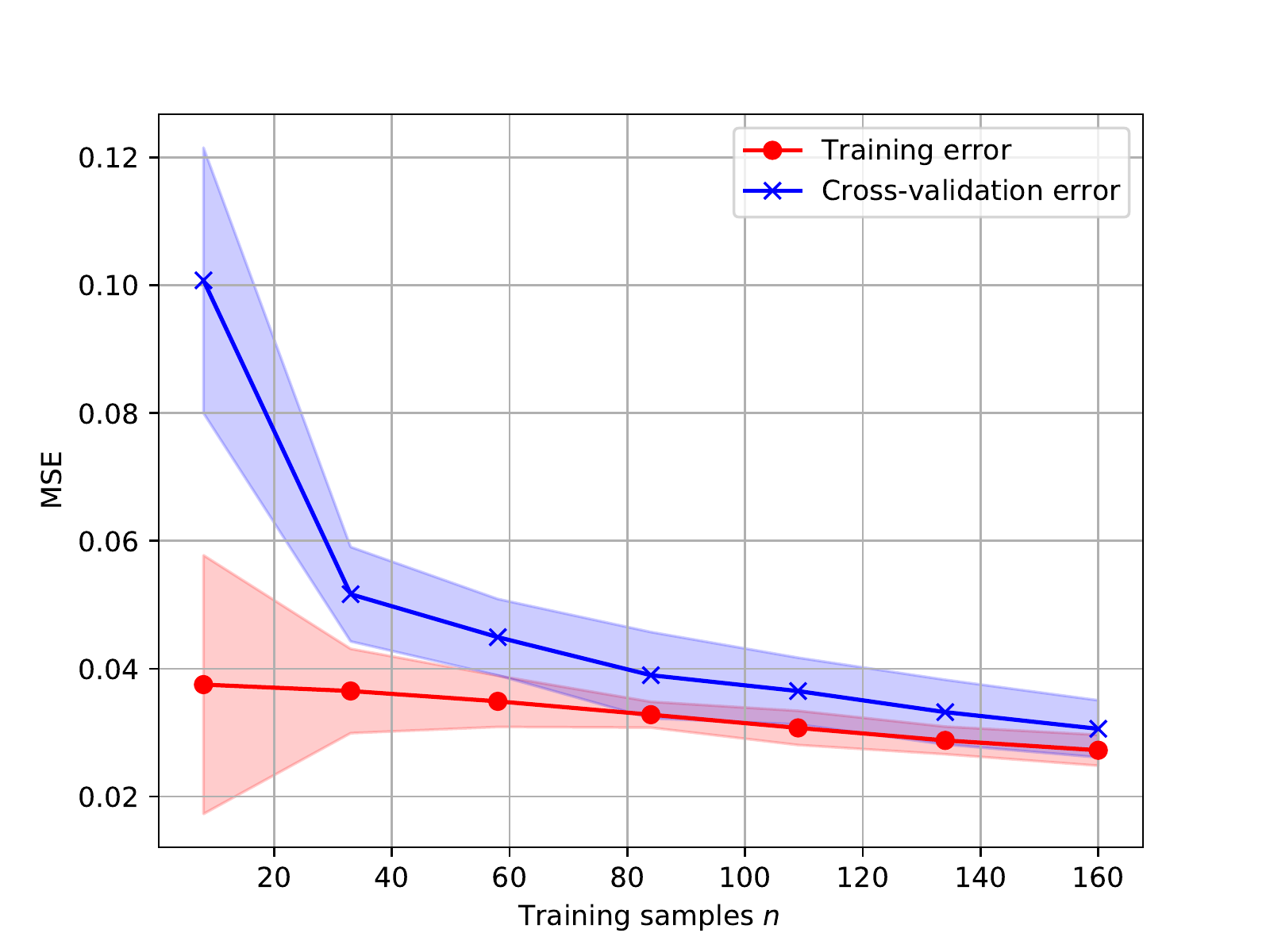}
	\caption{Learning curve based on a $10$-fold cross-validation with a testset size of $20\%$, where the prediction error is averaged over all timesteps. 
	We used a RBF kernel for the kPCA, the integrator and the pre-image KRR map. The parameters were chosen as before.}\label{fig:learning_curve_lkpca}
\end{figure}

The above $\ell$-kRR estimate is used to learn feature space integration and an approximate inverse transform of kPCA: Remember that a kPCA model trained from the data $\mathbf{x} = \{x_1,\hdots,x_m\} \subseteq \mathcal{X}$ can be used to compute the projections onto the principal axes of new data points $\mathbf{y} = \{y_1,\hdots,y_\ell\} \subseteq \mathcal{X}$. 
Let us denote these projections with $P_d \phi_k(y_i) \in \mathcal{F}_k,\,i=1,\hdots,\ell$. We are interested in an approximate backward transform which solves the pre-image problem, that is, learning a pre-image map $\Gamma:\, \mathcal{F}_k \rightarrow \mathcal{X}$ that approximates 
\begin{align}
 y_i \approx z_i = \Gamma P_d\phi_k(y_i), \, i=1,\hdots,\ell. 
\end{align}
This can be done by establishing a supervised learning model with $\ell$-kRR that tries to fit the training set $\mathbf{x}$ and its kPCA projections $P_d\phi_k(x_i),\,i=1,\hdots,m$, also cf. \cite{bakir2004learning}. 
The kRR problem that we define for determining the linear map $W:\, \mathcal{F}_k \rightarrow \mathcal{X}$ representing an approximation of $\Gamma$ takes a form analogue to \eqref{eqn:krr}
\begin{align}\label{eqn:krr_preimage}
 \min_W \frac{1}{2} \sum_{i=1}^m \|x_i - W\cdot\phi_k \big(P_d(\Phi_r(x_i))\big) \|^2 + \frac{\alpha}{2} \|W\|^2,
\end{align}
which can now be solved by the above low-rank variant. The $\ell$-kRR is also used to learn the feature space integration in a rather simple multi-step way by fitting $\nu \in \mathbb{N}$ feature vectors of reduced length $d$, corresponding to the time points $\{t, t+\Delta t, \hdots , t + (\nu-1) \Delta t\}$ and tagged by the respective field values, to the feature vectors associated with time point $t+ \nu \Delta t$. 
Fig. \ref{fig:kpca_mean_mag} and \ref{fig:kpca_mag_states} show mean magnetization and magnetization snapshots for the same fields as used by the NN-approach on the same test problem using the implementation from \cite{exl2020prediction}.
Fig.~\ref{fig:learning_curve_lkpca} shows the learning curve based on a $10$-fold cross-validation with a testset size of $20\%$, where the prediction error is averaged over all timesteps.   

\section{Conclusion}
Magnetization dynamics depending on external field is predicted by a two-stage scheme. First, training data obtained from micromagnetic numerical simulations are used to learn an embedding of the magnetization into 
a feature space with low-dimensional latent variable description. This is done by kernel methods or autoencoders, where in both cases an approximate inverse is learned. The induced feature space metric for 
the Gaussian kernel is locally isometric while simultaneously allowing for a low-rank approximation of the kernel matrix that enhances efficiency. We compare with a dense auto-/decoder embedding with no further norm 
preserving loss term or component scaling. In a second stage, the magnetization dynamics is learned for the respective low-dimensional representations in two different ways. Finally, we validate our approach by numerical examples. 
The kernel method approach is surprisingly accurate given its relatively simple underlying models based on explicitly known optimal solutions. However, several hyperparameters have to be adjusted.  
On the other hand, the autoencoder model naturally represents a data-dependent nonlinear dimensionality reduction that adapts to the specific problem and hence can achieve very high accuracy. 
However, training time for the overall NN approach exceeds that of the kernel methods by a factor of several magnitudes (in our test cases about $100$), 
in particular due to the rather complicated objective for the feature space integration scheme. However, predictions using the trained models become several magnitudes faster than direct computations. 
In addition, the low-rank approach represents an efficient way to train models which need large data sets, for instance the case when extended to stochastic LLG, which would be of great industrial interest. 

\section*{Acknowledgment}
We acknowledge financial support by the Austrian Science Foundation (FWF) via the projects "ROAM" under
grant No. P31140-N32 and the SFB "Complexity in PDEs" under grant No.
F65. The financial support by the Austrian Federal Ministry for Digital and Economic Affairs, the National 
Foundation for Research, Technology and Development and the Christian Doppler Research Association is gratefully acknowledged.
The authors acknowledge the University of Vienna
research platform MMM Mathematics - Magnetism - Materials. The computations were partly achieved by
using the Vienna Scientific Cluster (VSC) via the funded project No. 71140.

\bibliographystyle{abbrv}
\bibliography{bibref}

\end{document}